\documentclass[preprint,showpacs,preprintnumbers,amsmath,nofootinbib,amssymb]{revtex4}
\input psfig.sty
\usepackage{graphicx}% Include figurthebibliographye files
\usepackage{dcolumn}% Align table columns on decimal point
\usepackage{bm}% bold math

\begin{document}

\def\be{\begin{equation}}
\def\ee{\end{equation}}

%\preprint{APS/123-QED}

%\title{Gravitational lensing, supernovae and the time-dependence of
%energy}

\title{Nonextensivity in Geological Faults?}
%\title{Geophysical constraints on Tsallis nonextensive parameter}

%%%%%%%%%%%%%%%%%%%%%%%%%%%%%%%%%
\author{G. S. Fran\c{c}a} \email{george@dfte.ufrn.br}
\affiliation{Departamento de F\'\i sica, Universidade Federal do
R. G. do Norte, 59072-970, Natal, RN, Brasil}

\author{C. S. Vilar} \email{vilar@dfte.ufrn.br}
\affiliation{Departamento de F\'\i sica, Universidade Federal do
R. G. do Norte, 59072-970, Natal, RN, Brasil}

\author{R. Silva} \email{rsilva@uern.br,rsilva@on.br}
\affiliation{Observat\'orio Nacional, Rua Gal. Jos\'e Cristino 77,
20921-400 Rio de Janeiro - RJ, Brasil}
\affiliation{Universidade do Estado do Rio Grande do Norte,
59610-210, Mossor\'o, RN, Brasil}

\author{J. S. Alcaniz} \email{alcaniz@on.br}
\affiliation{Observat\'orio Nacional, Rua Gal. Jos\'e Cristino 77,
20921-400 Rio de Janeiro - RJ, Brasil}

\date{\today}% It is always \today, today,

\begin{abstract}
Geological fault systems, as the San Andreas fault (SAF) in USA, constitute typical examples of self-organizing systems in nature. In this paper, we have considered some geophysical properties of the SAF system to test the viability of the nonextensive models for earthquakes developed in [Phys. Rev. E {\bf 73}, 026102, 2006]. To this end, we have used 6188 earthquakes events ranging in the magnitude interval $2 < m < 8$ that were taken from the Network Earthquake International Center catalogs (NEIC, 2004-2006) and the Bulletin of the International Seismological Centre (ISC, 1964-2003). For values of the Tsallis nonextensive parameter $q \simeq 1.68$, it is shown that the energy distribution function deduced in above reference provides an excellent fit to the NEIC and ISC SAF data.  
\end{abstract}

% PACS,
                              %display SL99desired
\maketitle

\section{Introduction}

Earthquakes are among the most relevant paradigms of the so-called self-organized criticality, introduced by Bak, Tang and Wiesenfeld in Ref. \cite{bak87}. In the context of fault systems, this represents a complex spatio-temporal phenomenon which are related with the deformation and sudden rupture of some parts of the earth`s crust driven by convective motion in the mantle, i.e., the radiation of energy in the form of seismic waves. In particular, fault system as the San Andreas fault in California (USA) is one of the first example of self-organizing system in nature \cite{rundle02}. Despite the complexity of earthquakes, some of the known empirical laws are considerably simple as, for instance, the Omori law \cite{omori94} for temporal distribution of aftershocks and the Gutenberg-Richter law \cite{gut44} for relationship between frequency and magnitude. 

The physics of earthquakes has also been studied in the context of nonextensive formalisms \cite{T88,SL99}, where the very first investigation was done by Abe in Refs. \cite{abe03a,abe01,abe04}. Recently, a very interesting model for earthquake dynamics related to Tsallis nonentensive framework has also been proposed in Ref. \cite{oscar2004}. More recently, such a model was revisited by considering some new ingredients, namely, the correct definition for mean values in the context of Tsallis nonextensive statistics, and a new scale between the earthquake energy and the size of fragments, $\epsilon \propto r^3$; see Ref. \cite{nois2006} for details. In reality, by using the standard method of entropy maximization a new energy distribution function (EDF) was deduced, differing considerably from the one obtained in Ref. \cite{oscar2004}.   
 
In this paper, we test the viability of our approach \cite{nois2006} by using data taken from the Network Earthquake International Center catalogs (NEIC, 2004-2006) and the Bulletin of the International Seismological Centre (ISC, 1964-2003) for one of the best studied major fault zone in the world, i.e., the San Andreas Fault (SAF). As is well known, the SAF is one of the world longest and most active geological faults, reaching $\sim$15 kilometers deep with an extension of $\sim$1200 km and being about 20 million years old. This fault forms the boundary between the North American and Pacific plates and is classified as a right lateral strike-slip fault (transform boundary), although its movement also involves comparable amounts of reverse slip \cite{saf}. In the geological and geophysical context, a considerable number studies have been performed in order to better understand this complex system (see, e.g., \cite{saf1} and references therein). 

\section{Nonextensive description for earthquakes: a model}

In this Section, we recall the theoretical basis of our nonextensive earthquake model discussed in Ref. \cite{nois2006}. As well known, the Tsallis' framework generalizes the Botzmann-Gibbs statistics in what concerns the concept of entropy. Such formalism is based on the entropy formula given by

\begin{equation} \label{eq:1}
S_{q}=-\int p^q(\sigma)\ln_q p(\sigma) d\sigma.
\end{equation}

Here and hereafter, the Boltzmann constant is set
equal to unity for the sake of simplicity, $q$ is the nonextensive parameter and the $q$-logarithmic function above is defined by

\begin{equation}\label{eq:224}
\ln_q p\, =\, {(p^{1-q}-1)\over 1-q}, \,\,\,\,\,\,\,\,\, (p>0),
\end{equation}

which recovers the standard Boltzmann-Gibbs entropy $$S_1 = - \int p \ln p {d^{3}p}$$ in the limit $q\rightarrow 1$. In our model, $p(\sigma)$ stands for the probability of finding a fragment of relative surface $\sigma$, which is defined as a characteristic surface of the system. It is worth mentioning that most of the experimental evidence supporting Tsallis proposal are related to the power-law distribution associated with $S_q$ descripition of the classical $N$-body problem \cite{SPL98}\footnote{For a complete and updated list of references on Tsallis' entropy see http.tsallis.cat.cbpf.br/biblio.htm.}.

The fundamental idea of the model developed in Refs. \cite{oscar2004,nois2006} consists in the fact that the space between faults is filled with the residues of the breakage of the tectonic plates. In this regard, it was studied the influence of the size distribution of fragments on the energy distribution of earthquakes. The theoretical motivation follows from the fragmentation phenomena in the context of the geophysics systems \cite{englman87}. In this latter work, Englman {\it et al.} showed that the standard Botzmann-Gibbs formalism, although useful, cannot account for an important feature of fragmentation process, i.e., the presence of scaling in the size distribution of fragments, which is one of the main ingredients of our approach. Thus, a nonextensive formalism is not only justified in these earthquake  models but also necessary since the process of violent fractioning is very probably a nonextensive phenomenon, leading to long-range interactions among the parts of the object being fragmented (see, e.g., \cite{oscar2000}). 

In Ref. \cite{nois2006}, however, differently from \cite{oscar2004}, which assumes $\epsilon \sim r$, we used a new energy scale $\epsilon \sim r^3$. Thus, the proportionality between the released relative energy $\epsilon$ and $r^3$ ($r$ is the size of fragments) is now given by $\sigma - \sigma_q = (\epsilon/a)^{2/3}$, where $\sigma$ scales with $r^2$ and $a$ (the proportionality constant between $\epsilon$ and $r^3$) has dimension of volumetric energy density. In particular, this new scale is in full agreement with the standard theory of seismic rupture, the well-known seismic moment scaling with rupture length (see, for instance, \cite{thorne95}).

Following the standard method of maximization of the Tsallis entropy under the constrains of the normalization of the function and $q$-expectation value, we obtain (see \cite{nois2006} for details)

\begin{equation}\label{sigma}
p(\sigma)= \left[1-{(1-q)\over (2-q)}(\sigma - \sigma_q)\right]^{1 \over 1-q},
\end{equation}
which corresponds to the area distribution for the fragments of the fault, and the EDF of earthquakes reads as follows 

\begin{equation}\label{pe}
p(\varepsilon )d\varepsilon = \frac{C\varepsilon^{-1/3} d\varepsilon}{\left[1 + C'\varepsilon^{2/3}\right]^{1 \over q-1}},
\end{equation}
which has also a power-law form with $C$ and $C'$ given by
\begin{equation}
C={2\over 3a^{2/3}}\quad{\rm and}\quad C'=-{(1-q)\over(2-q)a^{2/3}}.
\end{equation}
In the above expression, the energy probability is written as $p(\varepsilon)=n(\varepsilon)/N$, where $n(\varepsilon)$ corresponds to the number of earthquakes with energy $\varepsilon$ and $N$ total number earthquakes.

\begin{figure}[t]
\vspace{.2in}
\centerline{\psfig{figure=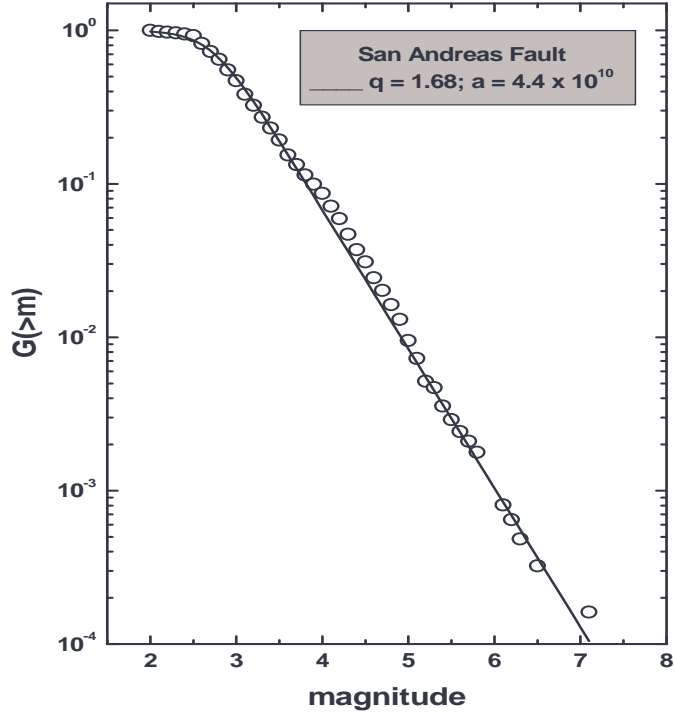,width=3.7truein,height=4.1truein}
\hskip 0.1in}
\caption{The relative cumulative number of earthquakes [Eq. (\ref{11})] as a function of the magnitude $m$. The data points correspond to 6188 earthquake events of San Andreas fault (lying in the interval $2 < m < 8$) from Network Earthquake International Center catalogs (NEIC, 2004-2006) and Bulletin of the International Seismological Centre (ISC, 1964-2003). The best-fit values for the parameters $q$ and $a$ are shown in the Panel. A summary of this and other recent analyses is also shown in Table I.}
\end{figure}

\section{Testing the model with experimental data}

In order to test the viability of the new EDF above derived [Eq. (\ref{pe})] we introduce the cumulative number of earthquakes, given by the integral
\cite{oscar2004}
\begin{equation}\label{inte}
{N_{\epsilon >}\over N} =\int _{\varepsilon}^\infty p(\varepsilon) d\varepsilon,
\end{equation}
or, equivalently,
\begin{equation}\label{inte1}
{N_{\epsilon >}\over N} =\int _{\varepsilon}^\infty \frac{C\varepsilon^{-1/3} d\varepsilon}{\left[1 + C'\varepsilon^{2/3}\right]^{1 \over q-1}},
\end{equation}
where $N_{\varepsilon >}$ is the number of earthquakes with energy larger than $\varepsilon$. By considering $m = 3^{-1} \log\varepsilon$ ($m$ stands for magnitude) it is possible to rewrite the above Eqs. as ($q\neq 1$)
\begin{eqnarray}\label{11}
\log (N_{>m}) & = & \log N + \left(\frac{2-q}{1-q}\right) \times \\ \nonumber & & \times \log \left[1 -
\left(\frac{1-q}{2-q}\right)\times \left({10^{2m}\over a^{2/3}}\right)\right],  
\end{eqnarray}
which, similarly to the modified Gutenberg-Ricther law (see, e.g., Refs. \cite{sornette} for more details), describes appropriately the energy distribution in a wider detectable range of magnitudes.

Figure 1 shows the main results of the present analysis. The relative cumulative number of earthquakes (defined as $G_{m>}=N_{m>}/N$) is shown as a function of the magnitude $m$. The data points, corresponding to earthquakes events (produced by the San Andreas fault) ranging in the 
interval $2 < m < 8$ with 6188 events, were taken from the Network Earthquake 
International Center catalogs for preliminary determinations (NEIC,
2004-2006) and Bulletin of the International Seismological Centre (ISC, 1964-2003). Note that we have used only the data from \emph{strike-slip movement} of the San Andreas fault. The best-fit parameters for this analysis are given by  $q = 1.68$ and $a = 4.4 \times 10^{10}$, which is very close to the values found for other geological faults at different parts of the globe. Note also that, similarly to our previous analyses, the nonextensive model of Ref. \cite{nois2006}, represented by Eqs. (\ref{sigma})-(\ref{11}), provide an excellent fit to the experimental data of the San Andreas fault. 

At this point we should compare the predictions of the model discussed above with the ones of Ref. \cite{oscar2004}. We note that a similar analysis to the latter model would provide the very same value for the nonextensive parameter, i.e., $q = 1.68$, although the energy density $a$ would differ by several orders of magnitude, $a = 4.3 \times 10^{-7}$. Naturally, this huge difference between the predictions of these nonextensive earthquake models could be used to check the viability of each model by comparing their predictions with experimental estimates for the energy density $a$. Note also that from most of the geophysical analyses performed so far values of $q \simeq 1.6 - 1.7$ seem to be \emph{universal}, in the sense that different data sets from different regions of the globe indicate a value for the nonextensive parameter lying in this interval. It is possible, however, that the predictions of the nonextensive earthquake models discussed in this paper be dependent on the fault mechanisms generating the earthquakes. A detailed analysis on the influence of the fault mechanisms (reverse, normal or strike-slip) on the predictions of these nonextensive models or, more specifically, on the values of the nonextensive parameter $q$ will appear in a forthcoming communication \cite{we2006}.

\begin{table}[h]
\caption{Limits to $q$ and $a$}
\begin{ruledtabular}
\begin{tabular}{lclc}
Fault& Ref. &\quad \quad $q$ & \quad $a$ \\
\hline \hline \\
California - USA & \cite{oscar2004} & \quad $1.65$& \quad $5.73 \times 10^{-6}$\\
Iberian Penisula - Spain & \cite{oscar2004} & \quad $1.64$& \quad $3.37 \times 10^{-6}$\\
Andaluc\'{\i}a - Spain & \cite{oscar2004} & \quad $1.60$& \quad $3.0 \times 10^{-6}$\\
Samambaia - Brazil & \cite{nois2006} & \quad $1.66$& \quad $1.8 \times 10^{10}$\\
New Madrid - USA & \cite{nois2006} & \quad $1.63$ & \quad $1.2 \times 10^{10}$\\
Anatolian - Turkey & \cite{nois2006} & \quad $1.71$ & \quad $2.8 \times 10^{10}$\\
San andreas - USA & This Paper & \quad $1.68$ & \quad $4.4 \times 10^{10}$\\
\end{tabular}
\end{ruledtabular}
\end{table}

\section{conclusion}

We have tested the viability of the earthquake model developed in Ref. \cite{nois2006} from the best studied major fault zone in the world, i.e., the San Andreas fault. By using 6188 earthquake events (in the interval $2 < m < 8$) taken from the Network Earthquake International Center catalogs and Bulletin of the International Seismological Centre we have shown, in agreement with other similar analyses, that for values of the nonextensive parameter of the order of $q = 1.6 - 1.7$ and $a \sim 10^{10}$, the model of Ref. \cite{nois2006} provides an excellent fit to the strike-slip movement data of the San Andreas fault (see Figure 1). We have also noted that although the predicted values for $q$ are very similar in our model and the model of Ref. \cite{oscar2004}, the value for the energy density $a$ differ by 17 orders of magnitude, which could be used to check the viability of both models (a summary of main results of the present and other recent analyses is shown in Table I). Finally, it is worth mentioning that the estimate for the nonextensive parameter from the San Andreas Fault data considered in this paper is in full agreement with the upper limit $q < 2$ obtained from several independent studies involving the Tsallis nonextensive framework \cite{newref}.

{\it Acknowledgments:} The authors thank the partial support by the Conselho Nacional de Desenvolvimento Cient\'{\i}fico e
Tecnol\'ogico (CNPq - Brazil). CSV is supported by CNPq/FAPERN. GSF is supported CNPq 309975/2003-4. JSA is supported by CNPq under Grants No. 307860/2004-3 and 475835/2004-2 and by Funda\c{c}\~ao de Amparo \`a Pesquisa do Estado do Rio de Janeiro (FAPERJ) No. E-26/171.251/2004.

\end{document}